\documentclass[prd,aps,preprint,tightenlines,onecolumn,
               superscriptaddress,nofootinbib,eqsecnum,
               preprintnumbers,longbibliography,floatfix,12pt]{revtex4-1}

\pdfoutput=1

\usepackage[T1]{fontenc}
\usepackage{amsmath,amssymb,amsthm,amsfonts}
\usepackage{mathrsfs,latexsym}
\usepackage{graphicx,tabularx,multirow}
\usepackage{caption}
\usepackage{subcaption}
\usepackage{enumerate,braket,xspace}
\usepackage{slashed,cancel}
\usepackage{url,xcolor,soul}
\usepackage{comment}
\usepackage{physics}

\usepackage[colorlinks=true,
            urlcolor=blue,
            anchorcolor=blue,
            citecolor=blue,
            filecolor=blue,
            linkcolor=blue,
            menucolor=blue,
            linktocpage=true,
            pdfproducer=medialab,
            pdfa=true]{hyperref}

\usepackage{cleveref}
\crefname{section}{Sec.}{Secs.}
\crefname{figure}{Fig.}{Figs.}
\crefname{equation}{Eq.}{Eqs.}
\crefname{appendix}{Appendix}{Appendices}

\newcommand{\gev}{\text{GeV}}

\definecolor{darkpink}{HTML}{FF69B4}

\usepackage[normalem]{ulem}

\newcommand{\bea}{\begin{eqnarray}\begin{aligned}}
\newcommand{\eea}{\end{aligned}\end{eqnarray}}
\newcommand{\gd}{g_d}

\usepackage{bm,epstopdf,natbib,hyperref,color,verbatim,multirow,bm,tikz,xcolor}
\hypersetup{colorlinks=true,urlcolor=blue,citecolor=blue,linkcolor=blue,menucolor=blue,anchorcolor=blue,filecolor=blue}
\everymath{\displaystyle}
\definecolor{lime}{HTML}{A6CE39}
\DeclareRobustCommand{\orcidicon}{
	\begin{tikzpicture}
	\draw[lime, fill=lime] (0,0) 
	circle [radius=0.16] 
	node[white] {{\fontfamily{qag}\selectfont \tiny ID}};
	\draw[white, fill=white] (-0.0625,0.095) 
	circle [radius=0.007];
	\end{tikzpicture}
	\hspace{-3mm}
}
\foreach \x in {A, ..., Z}{\expandafter\xdef\csname orcid\x\endcsname{\noexpand\href{https://orcid.org/\csname orcidauthor\x\endcsname}
			{\noexpand\orcidicon}}
}

\begin{document}
\preprint{}
	
\title{Dedicated Searches for Millicharged Particles at Intensity-Frontier Facilities: SpinQuest and SHiP}

\author{Leo Bailloeul}
\affiliation{Department of Physics, University of California, Davis, CA 95616, USA}
\author{Matthew Citron}
\affiliation{Department of Physics, University of California, Davis, CA 95616, USA}
\author{Yanou Cui}
\email{yanou.cui@ucr.edu}
\affiliation{Department of Physics, University of California, Riverside, CA 92521, USA}
\author{Saeid Foroughi-Abari}
\email{saeidf@physics.carleton.ca}
\affiliation{Department of Physics, Carleton University, Ottawa, Ontario K1S 5B6, Canada}
\author{Insung Hwang}
\affiliation{Department of Physics, Boston University, Boston, MA 02215, USA}
\author{Fengyi Li}
\affiliation{Department of Physics, University of California, Riverside, CA 92521, USA}
\author{Yu-Dai Tsai\hspace{-1mm}\orcidG{}}
\email{yt444@cornell.edu; yudaitsai.academic@gmail.com}
\affiliation{Los Alamos National Laboratory (LANL), Los Alamos, NM 87545, USA}
\author{Ming Xiong Liu}
\email{mliu@lanl.gov}
\affiliation{Los Alamos National Laboratory (LANL), Los Alamos, NM 87545, USA}
\author{Kranti Gunthoti}
\email{kranti@lanl.gov}
\affiliation{Los Alamos National Laboratory (LANL), Los Alamos, NM 87545, USA}
\author{Jae Hyeok Yoo}
\affiliation{Department of Physics, Korea University, Seoul, 02841, Korea}

\begin{abstract}
We conduct a dedicated study of searches for millicharged particles (mCPs) utilizing scintillator-based detectors at high-intensity fixed-target experiments, with particular focus on the SpinQuest and forthcoming Search for Hidden Particles experiment (SHiP) facilities. The analysis incorporates the three primary production mechanisms: meson decays, Drell-Yan processes, and proton bremsstrahlung. In particular, our updated analysis reveals that proton bremsstrahlung dominates the production rate in the sub-GeV mass regime. Detailed detector simulations and background evaluations are performed to obtain realistic sensitivity estimates. Our results demonstrate that future experiments located in the SpinQuest and SHiP facilities can achieve substantial improvements in discovery potential, enhancing sensitivity to the mCP charge parameter $\epsilon=q_{\chi}/e$ (with $q_\chi$ denoting the mCP electric charge) by up to two orders of magnitude relative to existing limits.

\end{abstract}

\maketitle

\tableofcontents

\section{Introduction}
Fractionally charged and millicharged particles (mCPs), exotic states with small electric charges, appear in many well-motivated extensions of the Standard Model (SM). They are closely connected to theoretical attempts that account for the observed quantization of electric charges, prediction of magnetic monopoles~\cite {Dirac:1931kp,Schwinger:1966nj}, and can serve as an indirect test of Grand Unified Theories (GUTs)~\cite{Pati:1973uk}.
Fractionally charged states can arise as characteristic low-energy consequences of string compactifications~\cite {Wen:1985qj}, and millicharged states can be effectively generated through kinetic mixing~\cite{Holdom:1985ag}.
Lately, the combination of accelerator searches and mCP cosmology has been used to probe early-universe reheating~\cite {Vogel:2013raa,Gan:2023jbs,Boddy:2024vgt}, in addition to the aforementioned high-energy theories, shedding new light on various aspects of fundamental physics.

A broad program to probe millicharged particles (mCPs) has emerged across many experimental frontiers.
Beyond conventional dark-matter searches—where experiments such as LZ~\cite{LZ:2022results,LZ:2025xkj} constrain small electromagnetic couplings of dark matter, neutrino facilities have provided strong limits and sensitivity projections for mCPs. The studies at LSND, MiniBooNE, MicroBooNE, SBND, DUNE~\cite{Magill:2018tbb}, FLArE~\cite{Kling:2022FLArE}, and ArgoNeuT~\cite{Harnik:2019zee,Acciarri:2020ArgoNeuTmCP} offer powerful constraints and sensitivity projections in complementary kinematic regimes. In collider experiments, several dedicated efforts are underway. Searches at the LHC include scintillator-based detectors such as milliQan~\cite{Ball:2016zrp,Alcott:2025rxn} and FORMOSA~\cite{Foroughi-Abari:2020qar,Citron:2025kcy}, while CMS has pursued direct mCP signatures in proton–proton collisions~\cite{CMS:2024eyx}. Additional fixed-target and beamline projects—FerMINI at Fermilab~\cite{Kelly:2018brz}, SUBMET at J-PARC~\cite{SUBMET}, and the LANSCE-mQ proposal at LANL~\cite{Tsai:2024wdh}, extend the reach to lower masses and smaller charges. A parallel set of efforts explores advanced detector technologies. Skipper-CDDs~\cite{Tiffenberg:2017Skipper} and superconducting-nanowire approaches~\cite{Hochberg:2019SNSPD} have enabled ultralow-threshold searches carried out by SENSEI~\cite{SENSEI:2024yyt,SENSEI:2025SNOLAB}, OSCURA~\cite{Oscura:2023Skipper}, and QROCODILE~\cite{QROCODILE:2025}.
Finally, mCPs and mCP dark matter may leave signatures outside laboratory settings, with potential impacts on cosmological probes such as CMB anisotropies~\cite{Dubovsky:2003yn}.

While most experimental projections in the literature have focused on mCP production via meson decays and parton-level processes, e.g., Drell-Yan, proton bremsstrahlung is shown to be the primary production mode at $\sim\gev$ scales at high-intensity fixed-target and collider experiments~\cite{Blumlein:2013cua,deNiverville:2016rqh,Foroughi-Abari:2021zbm, Batell:2020vqn,Blinov:2021say,Foroughi-Abari:2024xlj,Kling:2025udr}. The proton bremsstrahlung channel benefits from resonant enhancement via vector/scalar meson mixing and dominates in the forward region, offering enhanced detection rates compared to meson decay channels due to the broader angular distribution of the produced dark states in these decays.

In this work, we present comprehensive sensitivity projections for dedicated mCP searches at two high-intensity proton beam facilities: SpinQuest at the Fermilab Main Injector~\cite{Apyan:2022tsd} and SHiP at CERN Super Proton Synchrotron (SPS)~\cite{SHiP:2021nfo}.
Our analysis incorporates the proton bremsstrahlung channel alongside conventional meson decay modes, demonstrating substantial improvements in projected sensitivity across the MeV-to-GeV mass range.

The remainder of this paper is organized as follows. In~\cref{sec:model}, we review the theoretical framework and motivation for millicharged particles. Baseline detector setup and a summary of experimental facilities are presented in~\cref{sec:searches}. In~\cref{sec:production}, we describe the relevant production mechanisms and kinematic distributions of mCPs from high-energy proton beams, with particular emphasis on new results from the proton bremsstrahlung channel. Additional technical details related to this channel are provided in two appendices. We discuss potential background sources in~\cref{sec:backgrounds} and the projected sensitivities at the future experimental facilities are analyzed in~\cref{sec:sensitivity}. 
Finally, we present a discussion of the ongoing theoretical debate, comparing experimental sites and alternative configurations in~\cref{sec:discussion}.

\section{Millicharged Particles Model}\label{sec:model}

mCPs can arise through two primary mechanisms. In the minimal scenario (Sec.~\ref{sec:minimal}), the mCP directly couples to the SM hypercharge gauge group. Alternatively, the mCP can acquire a small electric charge through kinetic mixing with a dark photon (Sec.~\ref{sec:darkphoton}). Both frameworks lead to identical experimental signatures and detection strategies.

\subsection{Minimal mCP}\label{sec:minimal}

The minimal model for mCPs is described by the following Lagrangian:
\bea \label{eq:minimal MCP} \label{eq:minimal mcp lagrangian}
\mathcal{L} \supset i \overline{\chi} (\slashed{\partial} - i g' \epsilon \slashed{B} + m_\chi) \chi - \frac{1}{4} B_{\mu \nu} B^{\mu \nu}.
\eea

Here, $g'=e/\cos\theta_w$ is the gauge coupling of $U(1)_Y$, with $\theta_w$ as the weak mixing angle. $B_\mu$ is the SM hypercharge $U(1)_Y$ gauge boson. Once electroweak symmetry is broken, the hypercharge gauge field mixes according to $B_\mu = \cos\theta_w A_\mu - \sin\theta_w Z_\mu$, which endows $\chi$ with an electric charge parameter $\epsilon$.
Notably, the fractionally charged particles are a generic prediction of string compactification~\cite {Wen:1985qj}, and irrationally-charged particles could be a test of these compactification scenarios~\cite {Shiu:2013wxa}.

\subsection{Dark Photon mCP}\label{sec:darkphoton}

Beyond the minimal setup introduced above, mCPs may naturally arise in scenarios that include a dark photon. To illustrate this, we introduce a new particle $\chi$ that carries charge under an extra Abelian gauge group $U(1)_d$ with gauge boson $A'$, coupling strength $g_d$, and corresponding $\alpha_d \equiv g_d^2/4\pi$. The dark gauge field $A'$, which we refer to as the dark photon, can kinetically mix~\cite{Holdom:1985ag} with the SM hypercharge gauge field $B$ associated with $U(1)_Y$; for comprehensive discussions, see Refs.~\cite{Fabbrichesi:2020wbt,Agrawal:2021dbo}. The dynamics relevant for our discussion are encoded in the Lagrangian:
\bea
\label{eq:mcp_KM_L}
\mathcal{L} & \supset i \overline{\chi}(\slashed{\partial} - i \gd \slashed{A}' + m_\chi) \chi \\
& \quad - 
\frac{1}{4} B_{\mu \nu} B^{\mu \nu} - \frac{1}{4} A'_{\mu \nu}A'^{\mu \nu} + \frac{\kappa}{2 \cos \theta_w} B_{\mu\nu} A'^{\mu \nu},
\eea
where $A'_{\mu\nu}=\partial_\mu A'_\nu-\partial_\nu A'_\mu$ denotes the dark-photon field strength, $B_{\mu\nu}=\partial_\mu B_\nu-\partial_\nu B_\mu$ is the hypercharge field strength, and $\kappa$ quantifies the kinetic mixing between the $U(1)_Y$ and $U(1)_d$ sectors.

When the dark photon is massless and the $U(1)_d$ symmetry remains unbroken, one may eliminate the kinetic mixing, up to corrections of order $\mathcal{O}(\kappa^2)$, valid in the limit $\kappa \ll 1$, by redefining the dark gauge field as $A'_\mu \rightarrow A'_\mu + \kappa B_\mu/\cos\theta_w$~\cite{Holdom:1985ag}. Expressed in this diagonalized basis, the interaction terms take the form
\bea
\label{eq:chi_B_coupling}
    \mathcal{L} \supset g' \epsilon~\overline{\chi} \gamma^\mu \chi  B_{\mu}, \,\, \text{where $\epsilon = \frac{\kappa g_d}{e}$ },
\eea
with $g' = e/\cos\theta_w$ the hypercharge coupling. After electroweak symmetry breaking, the hypercharge gauge boson decomposes as $B_\mu {=} \cos\theta_w A_\mu {-} \sin\theta_w Z_\mu$, implying that the field $\chi$ acquires an effective electromagnetic coupling characterized by the millicharge parameter $\epsilon$.

The mCP can be detected via its interactions with SM particles, which is enabled by its coupling to the SM photons. Both theoretical frameworks (the minimal mCP model and the one with dark photon) yield almost identical experimental signatures in accelerator searches (unless, for example, there is background mCP dark matter~\cite{Berlin:2023gvx})  
The combination of cosmology and accelerator searches can potentially differentiate them and reveal the underlying fundamental particle theory~\cite{Gan:2023jbs}.
Throughout this paper, we parameterize mCP phenomenology in terms of the particle mass, $m_\chi$, and the effective fractional charge relative to the electron charge, $\epsilon$.

In addition to the minimal and dark-photon mCPs, there are also more complicated mCP models involving the Stueckelberg mechanism~\cite{Feldman:2007wj,Cheung:2007ut}. These models have more complex particle content, which may affect accelerator searches. In this paper, we will focus only on simple mCP models, but accelerator searches can be extended to study more complicated models.

\section{Detector setup and experimental facility}\label{sec:searches}

A high-energy beam impinging on a fixed target produces a large flux of mCPs, which can be detected with scintillator-based detectors. In this work, we adopt the nominal detector configuration that has been extensively studied in previous works~\cite{Haas:2014dda, Ball:2016zrp, Ball:2020dnx, Kelly:2018brz, Foroughi-Abari:2020qar, Citron:2025kcy, SUBMET}, consisting of $n$-layers of $18\times18 = 324$ scintillator bars optically coupled with photomultiplier tubes (PMTs).
We consider Eljen-200 scintillators~\cite{Eljen:EJ200} with dimensions $5$ $\times$ $5$ $\times$ $150$~cm$^3$, each read out by Hamamatsu R7725 PMT~\cite{Hamamatsu:R7725}.

mCPs traversing each bar deposit only a small amount of energy, resulting in signals in the PMTs that are comparable to single photoelectron pulses.
This forms the characteristic signature used to identify mCP events.
Since beam-produced mCPs are highly directional and time correlated with each beam spill, requiring coincident hits across multiple detector layers provides strong signal discrimination and significantly suppresses backgrounds.
A detailed description of signal simulation is provided in~\cref{sec:signal}.

In the remainder of this section, we present the two possible experimental sites where such a setup could be realized: SpinQuest at Fermilab, utilizing the $120$~GeV Main Injector proton beam, and SHiP at CERN SPS with a $400$~GeV proton beam.
By placing the detector at an appropriate place, along or near the beam line with sufficient shielding, it can enjoy the forward-peaked high flux of mCPs from the beam while SM particles are largely attenuated, either scattered out or losing their energy in the shielding and surrounding material, thereby reducing backgrounds and improving sensitivity to mCP signals.

\subsection{SpinQuest}

\begin{figure}
    \centering
    \includegraphics[width=1.0\linewidth]{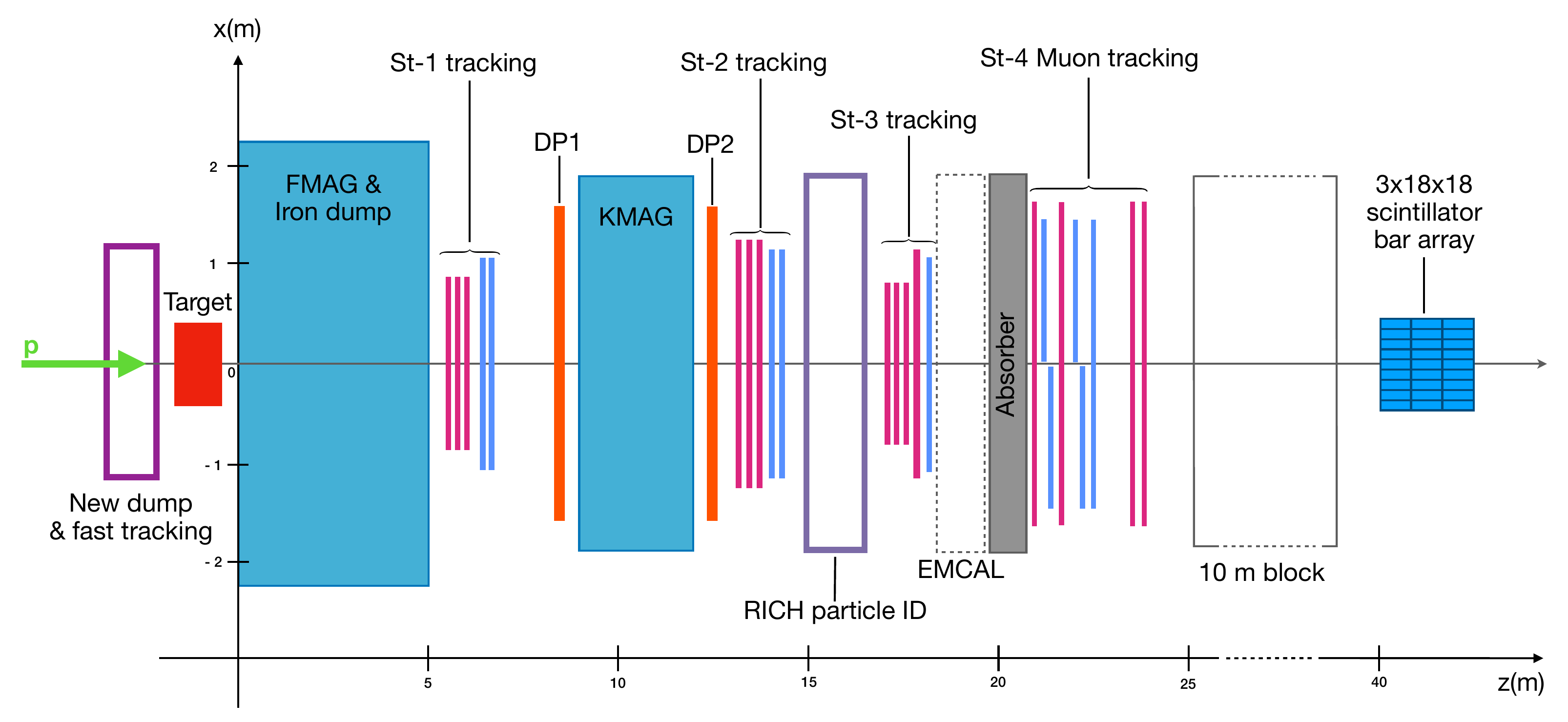}
    \caption{Schematic of the SpinQuest detector, modified from Ref.~\cite{Apyan:2022tsd}, showing the placement of a proposed 3-layer $18 \times 18$ scintillator bar mCP detector located approximately $40$ m downstream of the target.}
    \label{fig:DarkQuest_detector_geometry}
\end{figure}

SpinQuest is a fixed target spectrometer experiment at the Fermilab Accelerator Complex, which receives a high-intensity beam of 120 GeV protons from the Main Injector. SpinQuest studies the dimuon pairs from Drell–Yan interactions, produced when the beamline hits a polarized NH\(_3\) and ND\(_3\) target to study nucleon structure. DarkQuest and LongQuest are proposed upgrades to the SpinQuest program; DarkQuest proposes adding an electromagnetic calorimeter (ECAL) at $z\approx 19$ m downstream of the target, together with upgraded data acquisition systems. 
LongQuest proposes further enhancements: (1) a particle identification detector to discriminate between electrons, pions, protons, and kaons, and (2) a downstream detector at $\sim40$~m to search for more displaced signatures in a cleaner environment. 
The Main Injector beam delivers $\sim 4~$s spills every 60~s with $\sim 10^{12}$ protons per spill. Within each spill, the 53~MHz RF bunch structure corresponds to 18.8~ns spacing between microbunches. In terms of accumulated luminosity, DarkQuest phase I aims to collect $\sim 10^{18}$ protons on target (POT) in two years of running, while a future phase II could reach $\sim 10^{20}$ POT. In this work, we assume an accumulated POT of $10^{20}$ over a period of nominal phase II operation of SpinQuest. 
We consider an mCP detector located 40 meters downstream of the target, located behind the 10-meter iron block shown in Fig.~\ref{fig:DarkQuest_detector_geometry}. In this schematic, each illustrated bar represents two scintillator bars for a detector height of 18 bars (or $\sim 90$ cm) centered about the z-axis. The 5 m magnetized iron block (FMAG), serves both as a beam dump and magnetic sweeper, while the downstream open-aperture KMAG further deflects SM radiation. Together with its compact geometry, high beam energy, and thick hadron absorber, this configuration provides substantial shielding from beam-related backgrounds and makes SpinQuest an attractive, cost-effective setting for probing a broad range of dark-sector particles.

\subsection{SHiP} 

SHiP is a beam dump experiment that will be installed at the ECN3 facility at the SPS, located 8 meters below ground. 
The SPS facility is a synchrotron particle accelerator at CERN that delivers a $400$~GeV high-intensity proton beam.
The SHiP beamline delivers spills of duration of $\sim$ 1~s, each containing $4\times 10^{13}$ protons at 400 GeV with a non-uniform spill structure. 
One year of run time at SPS is equivalent to $10^6$ spills, accumulating  $4\times 10^{19}$ POT/year.
Throughout this work, we assume $5$ years of nominal SPS operation, corresponding to a total of $2\times 10^{20}$ POT.
The proton target consists of a dense mix of tungsten and molybdenum, followed by the hadron absorber and the magnetic deflector, called the muon shield, with the purpose of suppressing muons and neutrinos from pion and kaon decays before they reach the detector complex.

\begin{figure}
    \centering
    \includegraphics[width=1.0\linewidth]{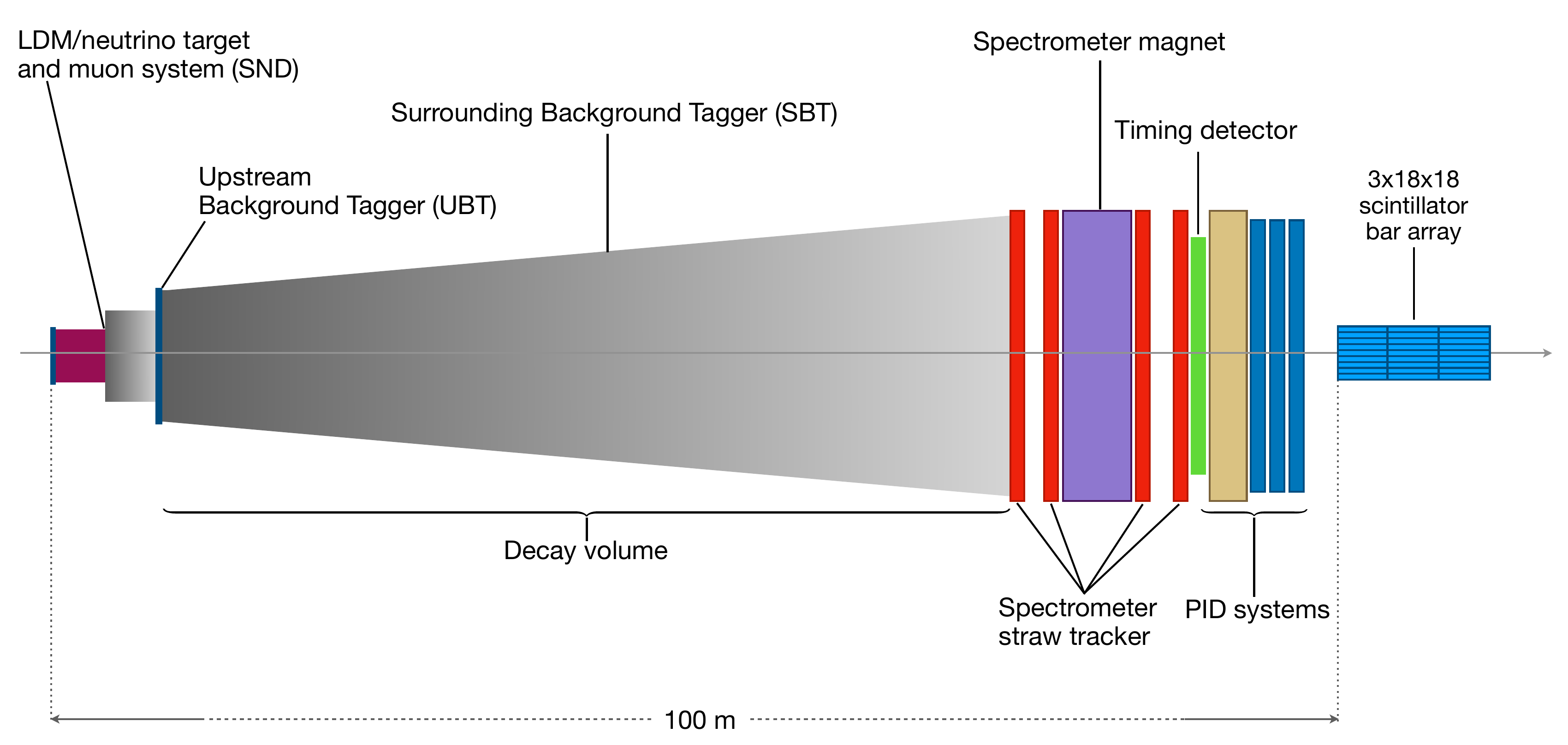}
    \caption{Schematic layout of the SHiP, adapted from Ref.~\cite{Albanese:2878604}, with the proposed three-layer $18 \times 18$ scintillator bar mCP detector setup located approximately $100$ m downstream of the target.}
    \label{fig:SHiP_detector_geometry}
\end{figure}

In this study, we consider an additional mCP detector located $100$ m downstream of the target, placed after the end of the decay volume and between the straw-tracker stations shown in Fig.~\ref{fig:SHiP_detector_geometry}.
In this configuration, the SHiP absorbers and muon shield reject beam-related backgrounds, the surrounding facility material provides additional shielding, and the tracking, timing, and PID capabilities of the SHiP spectrometer can further reduce residual backgrounds.

\section{mCP Production}\label{sec:production}
In this section, we discuss several production mechanisms of mCPs,  emphasizing their kinematic features and relative contributions, and provide a detailed description of the simulation procedure.
At sub-GeV to GeV masses, mCPs can be produced in proton collisions either directly or via the decay of secondary mesons.
Neutral meson decays provide an important source, including decays of pseudo-scalar mesons $\pi^{0},\eta \to \gamma \chi \overline{\chi}$ and decays of vector mesons $\mathfrak{m} \to \chi\overline{\chi}$ where we considered $\mathfrak{m} = \rho, \omega, \phi, J/\psi, \Upsilon$.
Beyond meson decays, we also include contributions from the Drell–Yan process and proton bremsstrahlung. 
Among all channels, proton bremsstrahlung dominates forward production for high-energy proton beams because of the collinear enhancement. As a direct production mode, proton bremsstrahlung also involves mixing with vector mesons, which resonantly enhance the mCP production rate. For heavier mCPs with masses well above the nucleon scale, the Drell–Yan process becomes particularly important. 
 
\subsection{Meson Decay}
One of the main contributions to mCP production at fixed target facilities is the decay of neutral pseudo-scalar mesons, $\pi^{0},\,\eta \to \gamma \chi\overline{\chi}$.
In the case of a massless dark photon mediator, the total number of mCPs produced in pseudo-scalar meson decays is given by 
\cite{Kelly:2018brz}
\begin{equation} \label{eq:3-body meson}
    N_{\chi} \simeq 2 N_{\mathfrak{m}}  \epsilon^{2}\alpha_{\text{EM}} \text{Br}(\mathfrak{m}\to\gamma\gamma) \times I^{(3)} \qty(\frac{m_{\chi}^{2}}{m^{2}_{\mathfrak{m}}}).
\end{equation}
Here, $ N_{\mathfrak{m}}$ is the number of meson $\mathfrak{m}$ produced and $I^{(3)}$ is a phase space integral given by
\begin{equation}
    I^{(3)}(x) = \frac{2}{3\pi} \int_{4x}^{1} dz \, \sqrt{1-\frac{4x}{z}} \frac{(1-z)^{3}(2x+z)}{z^{2}}.
\end{equation}
The angular distributions of these decays are simulated according to  
\begin{align}
&\frac{d\text{Br}(\mathfrak{m} \!\to\! \gamma\chi\chi)}{ds \, d\!\cos\theta}
 = \frac{\epsilon^2 \alpha_{\rm EM}}{4 \pi s} \qty\bigg(1\!-\!\frac{s}{m_{\mathfrak{m}}^2})^3 \qty\bigg(1-\frac{4m_\chi^2}{s})^{\frac12}   \qty\bigg[2-\qty\bigg(1-\frac{4m_\chi^2}{s} ) \sin^2 \theta]  \times \text{Br}(\mathfrak{m} \!\to \!\gamma\gamma),
\end{align}
where $s = (p_\chi + p_{\bar\chi})^2$  and $\theta$ denotes the polar angle of $\chi$ evaluated in the $V^{*}$ rest frame, with the $z$-axis chosen to follow the boost direction of $V^{*}$~\cite{Jodlowski:2019ycu}.
For higher mCP masses, direct decay of vector mesons to a pair of mCPs, $\mathfrak{m}\rightarrow \chi \bar{\chi}$ can contribute to the total mCP yield.
The number of mCPs produced in this channel can be approximated as
\begin{equation} \label{eq:2-body meson}
N_\chi \simeq 2 N_{\mathfrak{m}}\epsilon^2 \mathrm{Br}\left(\mathfrak{m} \rightarrow e^{+} e^{-}\right) \times I^{(2)}\left(\frac{m_\chi^2}{m_{\mathfrak{m}}^2}, \frac{m_e^2}{m_{\mathfrak{m}}^2}\right),
\end{equation}
where
\begin{equation}
I^{(2)}(x, y)=\frac{(1+2 x) \sqrt{1-4 x}}{(1+2 y) \sqrt{1-4 y}},
\end{equation}
accounts for the phase space differences between the decay into mCPs and the decay into electrons.
The meson yields per POT, and the meson spectra are simulated using \texttt{Pythia8}~\cite{pythia8.3, Sjostrand:2006za}.
We find that each proton on target produces $N_{\pi^{0}} = 4.7(7.5)$,  $N_{\eta} = 0.53(0.85)$, $N_{\rho} = 0.61(1.0)$, $N_{\omega} = 0.61(1.0)$, $N_{\phi} = 0.022(0.041)$, $N_{J/\psi} = 4.1\times 10^{-5}(8.3\times 10^{-5})$, and $N_{\Upsilon} = 2.5\times10^{-9}(5.5\times10^{-9})$\footnote{As $\Upsilon$ production at SpinQuest is too rare to simulate, we follow the approach made in \cite{Kelly:2018brz}.} for SpinQuest (SHiP), which are consistent with previous studies~\cite{Kelly:2018brz, Magill:2018tbb, Berlin:2018pwi, Harnik:2019zee, Coloma:2023adi, SHiP:2020noy}.
We define $N_{\chi}^{\text{Meson}}$ as the sum of (\ref{eq:3-body meson}) and (\ref{eq:2-body meson}) over all meson decay channels rescaled by the fraction of mCPs whose trajectories, sampled from the simulated angular distribution of each channel, intersect the detector's front face.
We find that $\mathcal{O}(10^{-2})$ of the total produced mCPs pass through the detector at SpinQuest and that $\mathcal{O}(10^{-3}) $ of the total produced mCPs pass through the detector at SHiP.

\subsection{Drell-Yan} \label{subsec:dy}
The production of mCPs via Drell-Yan process is estimated using \texttt{MadGraph5}~\cite{Alwall:2014hca}, implementing the Minimal mCP model described in \ref{sec:minimal} with \texttt{FeynRules}~\cite{Alloul:2013bka}.
We generate Drell-Yan events at leading order with the parton distribution function \texttt{nCTEQ15\_56\_26}~\cite{Kovarik:2015cma} for an iron target, and obtain the proton-iron cross section by rescaling the resulting cross section by the atomic mass $A$~\cite{Batell:2020vqn}.
The detector acceptance is estimated from the simulated angular distribution of mCPs, and the number of mCPs reaching the detector, $N_{\chi}^{\text{DY}}$, is computed by using the luminosity in Eq. (\ref{eq:luminosity}).
For $m_{\chi} \sim 1 $ GeV and $N_{\text{POT}} = 1$, we find $N_{\chi}^{\text{DY}} \approx 1.9 \times 10^{-9} \epsilon^2$ at SpinQuest and $N_{\chi}^{\text{DY}} \approx 5.3\times 10^{-9}\epsilon^2$ at SHiP.

\subsection{Proton Bremsstrahlung}
\label{sec:pbrem}

\begin{figure}[t]
\centering
\includegraphics[width=0.49\textwidth]{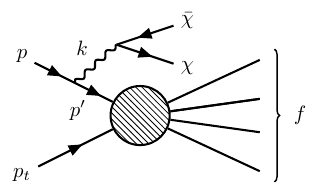}
\caption{Pair production of \(\chi\bar{\chi}\) via initial state radiation in a generic non–diffractive scattering event.}
\label{fig:FeymCP}
\vspace{-0.5cm}
\end{figure}

The proton bremsstrahlung process can be characterized with the initial– and final–state radiation and collective effects in the underlying hadronic collision. In \cite{Foroughi-Abari:2021zbm}, several approaches to proton bremsstrahlung were studied, including initial-state radiation (ISR) and final-state radiation in quasi-elastic scattering, the hadronic Weizsäcker–Williams (WW) approximation for quasi-elastic scattering, and an approach following Altarelli-Parisi to ISR in inelastic scattering referred to as the quasi-real approximation (QRA). Among these, the QRA approach was found to yield the dominant contribution to the total production rate. This is because, in quasi-elastic scattering, interference between ISR and FSR significantly suppresses the rate. Therefore, we focus on non-single diffractive events where both beam and target protons dissociate, as illustrated in Fig.~\ref{fig:FeymCP}. In such events, interference is absent and ISR dominates. 

More recently, modeling of proton bremsstrahlung has been improved~\cite{Foroughi-Abari:2024xlj,Kling:2025udr}, focusing on initial state radiation, refining the relevant splitting functions, and incorporating more accurate form factor parameterizations. These recent improvements provide a practical tool for predicting dark state production via proton bremsstrahlung. Nevertheless, further theoretical and experimental work is needed to refine the QRA formalism and reduce uncertainties that can significantly impact experimental sensitivity. Recently, a data-driven framework~\cite{Allison:2025mom} has also been proposed that determines the mCP production rates and kinematics directly from observed dilepton events, bypassing the need for model-dependent theoretical inputs.

A number of studies have explored the production of mCP pairs via proton bremsstrahlung, motivated in part by the relevance of this channel for atmospheric mCP fluxes. This process has been examined using quasi-elastic splitting functions~\cite{Du:2022hms} and, in other work, using the modified Weizsäcker–Williams (WW) approximation for dark photons~\cite{Blumlein:2013cua, Wu:2024iqm}. However, as shown in Refs.~\cite{Foroughi-Abari:2021zbm, Foroughi-Abari:2024xlj}, the modified WW kernel generically overestimates bremsstrahlung rates in inelastic collisions, where the QRA approach offers a more reliable description. More recently, Ref.~\cite{Kling:2025udr} studied mCP production in proton–proton collisions by relating it to dark photon production using an effective QRA splitting function with Dawson corrected polarization sum~\cite{Foroughi-Abari:2024xlj} and updated nucleon form factors. However, it remains unclear whether applying this correction preserves gauge invariance for resonant pair production mediated by a virtual photon, as discussed in detail in~\cref{app:phase_space_decompos}. 

In this work, we improve upon previous studies by deriving the ISR splitting probability for mCP pair production directly using the QRA formalism. Unlike earlier approaches, we do not rely on the 2 to 3 quasi-elastic process or the modified WW approximation. In our main analysis, we also do not use a factorized phase space decomposition, in which a massive photon is first produced via proton bremsstrahlung and subsequently decays into a pair. Instead, our computation treats the internal virtual photon conversion to a millicharged pair directly, ensuring consistency with gauge invariance.

The process of interest is the direct production of mCPs in proton–target collisions. We consider the forward production of mCP pairs via proton ISR of a virtual photon that subsequently converts to a \(\chi\bar\chi\) pair:
\begin{equation}\label{eq:brem-scattering}
p(p)+p(p_t)\ \rightarrow\ \chi(p_\chi)+\bar{\chi}(p_{\bar{\chi}})+f(p_f),
\end{equation}
where \(f\) denotes the inclusive hadronic final state. The virtual photon carries momentum \(k = (p_\chi+p_{\bar{\chi}})\) with invariant mass squared \(k^2 \ge 4m_\chi^2\). Our focus here is on ISR in proton–proton scattering, as it is well defined given specified initial states. In the QRA, or the on-shell approach, the intermediate proton $p'$ is treated as nearly on-shell, allowing the differential ISR cross section to factorize into a calculable splitting probability and the underlying non-single-diffractive proton–proton cross section, the latter of which can be obtained from experimental data. After integrating over the hadronic final state \(f\), the ISR cross section factorizes as
\begin{align}
    d\sigma^{pp\rightarrow \chi\bar\chi\, f}(s)
    \ \approx\ d\mathcal{P}_{p \rightarrow p^{\prime}\chi\bar\chi}\ \times\ \sigma_{pp}^{\rm NSD}(s^{\prime})\, ,
    \label{eq:factorization}
\end{align}
where the non–single diffractive (NSD) cross section, parametrized as
\begin{equation}
\sigma_{pp}^{\rm NSD}(s) \;=\; 1.76\ +\ 19.8\left(\frac{s}{\mathrm{GeV}^2}\right)^{0.057}\ \ \ \text{mb}\,,
\label{eq:sigmaNSD}
\end{equation}
is evaluated at a reduced center of mass energy $s^\prime \simeq2m_p(E_p{-}k^0)$, where $E_p$ is the beam energy, accounting for the fraction of beam energy carried away by the radiated pair. 

The differential splitting probability in Eq.~\eqref{eq:factorization} is given by
\begin{align}
d\mathcal{P}_{p \rightarrow p^{\prime}\chi\bar\chi}
&=
\frac{\left|\overline{\mathcal{M}^{\,p\rightarrow p^\prime \chi \bar{\chi}}}\right|^2}{\big((p{-}k)^2{-}m_p^2\big)^2}\,
\frac{|\vec p -\vec k|}{|\vec p|}\,
\frac{d^3p_{\chi}}{2E_{\chi}(2\pi)^3}\,
\frac{d^3p_{\bar{\chi}}}{2E_{\bar{\chi}}(2\pi)^3}
,
\label{eq:master}
\end{align}
where the expression for the squared matrix element of the sub-process is given in Eq.~\eqref{eqn:QRA_XS}, and the details are presented in~\cref{app:pBrem}. Equation \eqref{eq:master} is the splitting function used in our analysis.

This formalism can be readily generalized to the case of a proton impinging on a nuclear target, as relevant for beam-dump experiments, by replacing $\sigma_{pp}^{\rm NSD}(s)$ with an inclusive $pA$ cross section that excludes target-single-diffractive (TSD) configurations, thereby avoiding ISR/FSR interference. A fit to data yields~\cite{Carvalho:2003pza,Carroll:1978hc}:
\begin{equation}
\sigma_{pA}(s)=\sigma_{\mathrm{inel}} - \sigma_{\mathrm{TSD}}
= 43.55~\mathrm{mb}\, A^{0.7111}\ -\ 3.84~\mathrm{mb}\, A^{0.35}\,,
\label{eq:pA-fit-b}
\end{equation}
which is only weakly dependent on \(s\) for the \(\sqrt{s}\lesssim 30~\mathrm{GeV}\) regime relevant to proton beam energies at SpinQuest and SHiP. In practice, we treat it as approximately constant.

The validity of the factorized QRA approximation in Eqs.~\eqref{eq:factorization}–\eqref{eq:master} for the ISR process relies on specific kinematic conditions: the off-shell momentum of the intermediate proton and the transverse momentum of emission must be small compared to the hard scattering scale. Also, the radiated pair, the beam proton, and the intermediate proton should all be ultrarelativistic. The first condition is enforced by introducing an off-shell form factor at the $pp^\prime \gamma^\star$ vertex to suppress contributions from large virtualities. 
In addition, the proton’s electromagnetic form factors are introduced to model its internal structure in processes involving the coherent emission of a virtual photon with a timelike momentum. We implement a form factor in \eqref{eq:master} that can be parametrized as 
\begin{equation}\label{eq:formfactors}
    F(k^2, p^{\prime 2}) = \mathcal{K}_{pp^\prime \gamma^\star} (p^{\prime 2}) \times F(k^2),
\end{equation}
where $F(k^2)$ denotes the proton electromagnetic form factor, while
$\mathcal{K}_{pp^\prime \gamma^\star} (p^{\prime 2})$  encodes the off-shell dependence of the intermediate proton line and can be modeled via a simple dipole form \cite{Foroughi-Abari:2021zbm},
\begin{equation}
{\cal K}_{pp^\prime \gamma^\star} (p^{\prime 2}) = \frac{1}{1+(p'^2{-}m_p^2)^2/\Lambda_p^4},
\end{equation}
with a hadronic scale $\Lambda_p\sim \gev$ controlling the level of off-shell contributions. Varying this parameter between 1.0 and 2.0~$\gev$ provides an estimate of the uncertainty.

\begin{figure}[t]
  \centering
  \includegraphics[width=0.49\columnwidth]{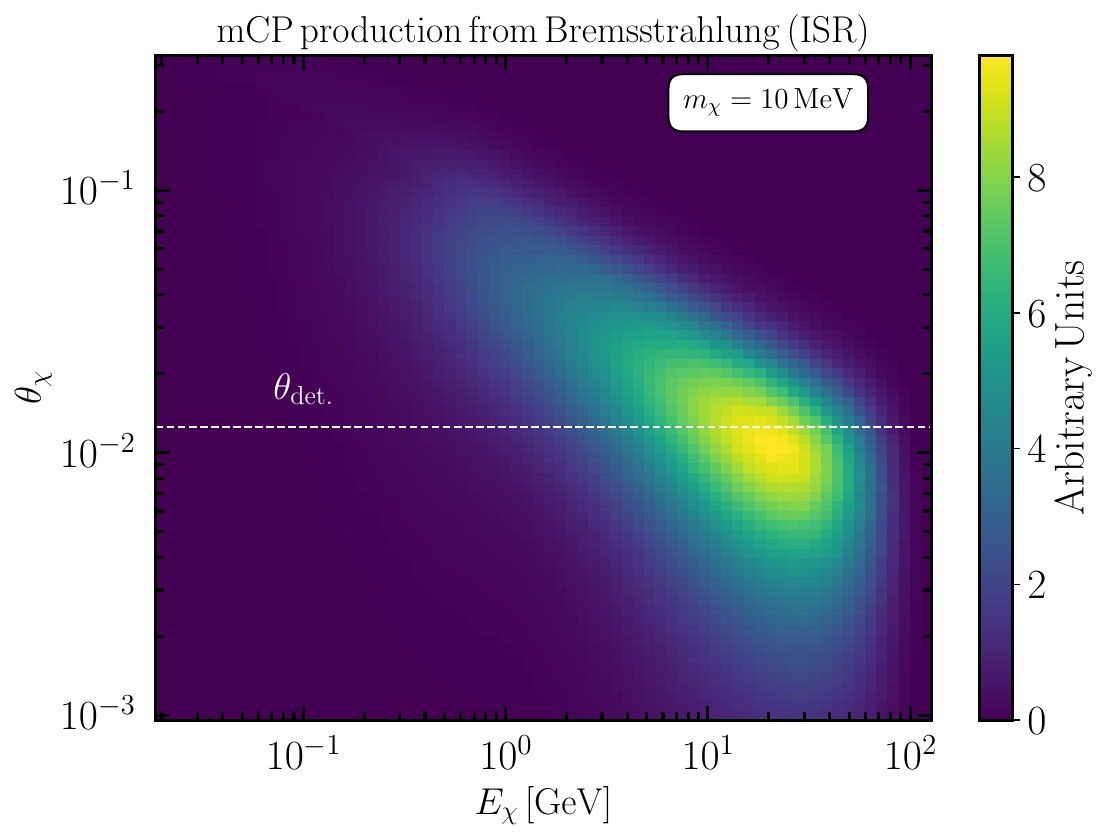}
  \hfill
  \includegraphics[width=0.49\columnwidth]{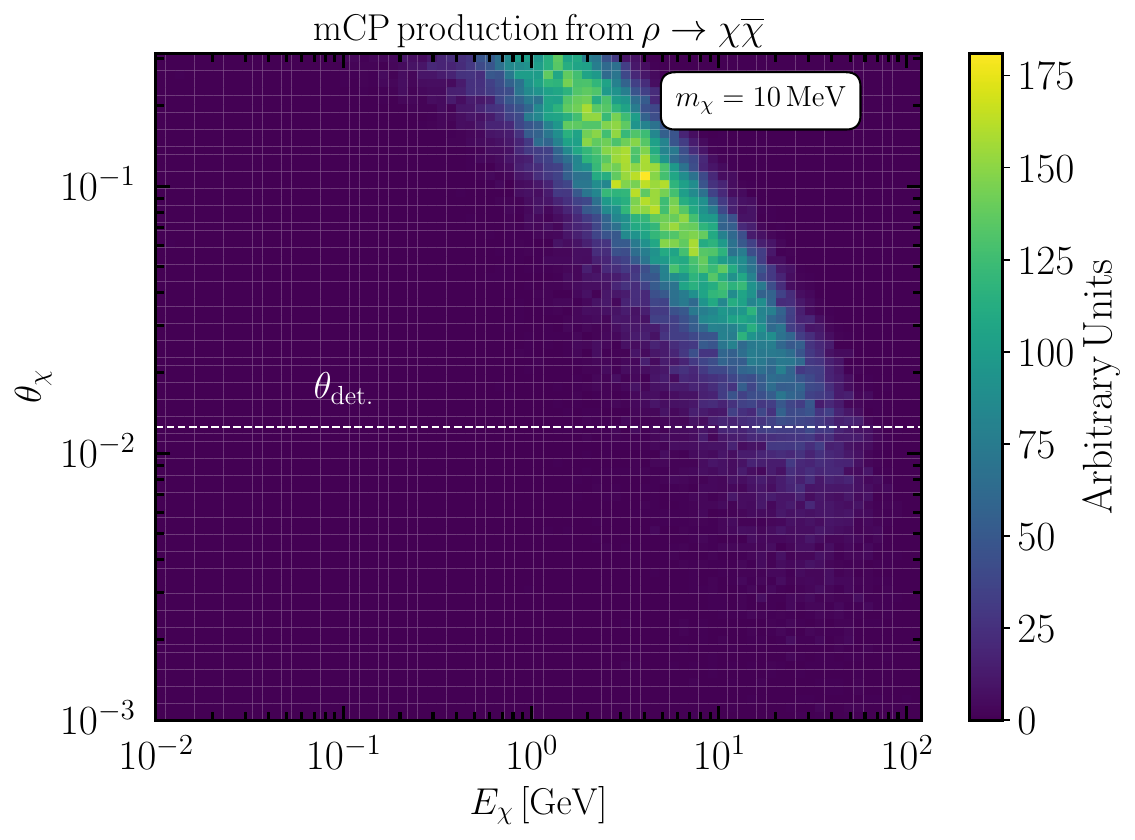}
  \caption{Distributions of mCP angle ($\theta_\chi$) relative to the beam axis and momentum ($p_\chi$) for $m_\chi = 10\,\mathrm{MeV}$, shown for the two main production mechanisms: proton bremsstrahlung (left) and $\rho$ meson decay (right).}
  \label{fig:distmCP}
\end{figure}

The electromagnetic structure of the proton induces resonant enhancement through mixing between the intermediate photon and the SM vector mesons $\rho$ and $\omega$. Recent analytical and phenomenological evaluations of the nucleon form factors, including the $\phi$ meson resonance, have been presented in Refs.~\cite{Foroughi-Abari:2024xlj, Adamuscin:2016rer}, and more recently, various parametrizations with good fits to data have been studied in~\cite{Kling:2025udr}. In this work, we adopt a simplified electromagnetic form factor to account for the mixing with $\omega$ and $\rho$ resonances, within the framework of the extended vector meson dominance (eVMD) model~\cite{Foroughi-Abari:2021zbm,deNiverville:2016rqh, Faessler:2009tn}. While less detailed than the aforementioned studies, it captures the essential physics. We retain only the Dirac form factor here, which has relatively small uncertainties for $k^2 \lesssim 1~\rm{GeV^2}$. The full matrix element, including Dirac, Pauli, and interference contributions, is provided in~\cref{app:pBrem}, where we also compare different form factor choices and estimate the associated uncertainties.

Employing the factorized ISR cross section~\cref{eq:factorization} and the derived splitting probability in~\cref{eq:master}, we estimate the total proton bremsstrahlung production cross section for millicharged pairs as 
\begin{equation}
\sigma_{\text{brem}} = \int_{p_\chi^{\min}}^{p_\chi^{\max}} \int_{\theta_\chi^{\min}}^{\theta_\chi^{\max}} dp_\chi\, d\theta_\chi \; \frac{d\mathcal{P}_{p \rightarrow p^{\prime}\chi\bar\chi}\ }{dp_\chi\,d\theta_\chi} \sigma_{pA}(s^{\prime}),
\end{equation}
scaling as $\epsilon^2$. 
The expected number of mCP hits from bremsstrahlung is 
\begin{equation} \label{eq:luminosity}
    N^{\text{Brem}}_{\chi} =   N_{\text{POT}}\left(\frac{\rho L }{A}N_A\right)  \times\sigma_{\text{prod}} ,
\end{equation}
where $\sigma_{\text{prod}}$ includes detector acceptance, $N_{\text{POT}}$ protons on target, $\rho$ the target density, $L = \lambda_{\text{int}}$ the nuclear interaction length in target, and $A$ the atomic mass. 

To obtain the mCP production rate, we perform a \texttt{vegas} adaptive Monte-Carlo integration algorithm~\cite{peter_lepage_2020_3897199} of~\cref{eq:factorization}, employing the cross section in~\cref{eq:pA-fit-b} and the form factors in~\cref{eq:formfactors} with kinematics specified in~\cref{app:pBrem}. The predicted bremsstrahlung rates are compared to other production modes for the SpinQuest and SHiP configurations, as shown in~\cref{fig:mcp_production}. 

We find that the total mCP production rate exhibits a characteristic rise at low masses, mirroring the $1/m_\chi^2$ scaling behavior of bremsstrahlung in QED-like processes~\cite{Freytsis:2009bh,Bjorken:2009mm,Izaguirre:2014bca,Gninenko:2018ter,Zhou:2024aeu,Blinov:2025aha}. This scaling arises from the interplay between the virtual photon propagator, the emission amplitude, and the phase space of the mCP pair. In~\cref{app:phase_space_decompos}, we examine this behavior in detail and show that the factorized expression in~\cref{eq:brem_mass_scaling} reproduces the same low-mass scaling and agrees with the full internal conversion calculation in~\cref{eq:master} when using the massive photon QRA splitting function of Ref.~\cite{Foroughi-Abari:2021zbm}, while differing from the result of Ref.~\cite{Kling:2025udr} due to a distinct choice of polarization sum at the splitting vertex, which suppresses the low-mass rise. Further discussion and derivations are provided in~\cref{app:phase_space_decompos}.

The bremsstrahlung production mode yields a higher rate of detectable mCPs compared to meson decays, primarily due to its strong collinear enhancement. In proton bremsstrahlung, the produced mCPs have a sharply forward-peaked distribution, closely aligned with the incoming proton beam direction. This results in a large fraction of the produced mCPs falling within the small angular acceptance of forward detectors, e.g., for $\theta_{\rm det} \lesssim 10\,\mathrm{mrad}$ region shown in Fig.~\ref{fig:distmCP}. In contrast, meson decays produce mCPs that are more isotropic in the meson rest frame, and although the mesons are forward boosted, the resulting angular spread of the mCPs is still broader than in the bremsstrahlung case. As a result, fewer mCPs from meson decays reach the detector.

\subsection{Detector Signal} \label{sec:signal}
\begin{figure}[!htb]
    \centering

    \begin{subfigure}{\linewidth}
        \centering
        \includegraphics[width=0.82\linewidth]{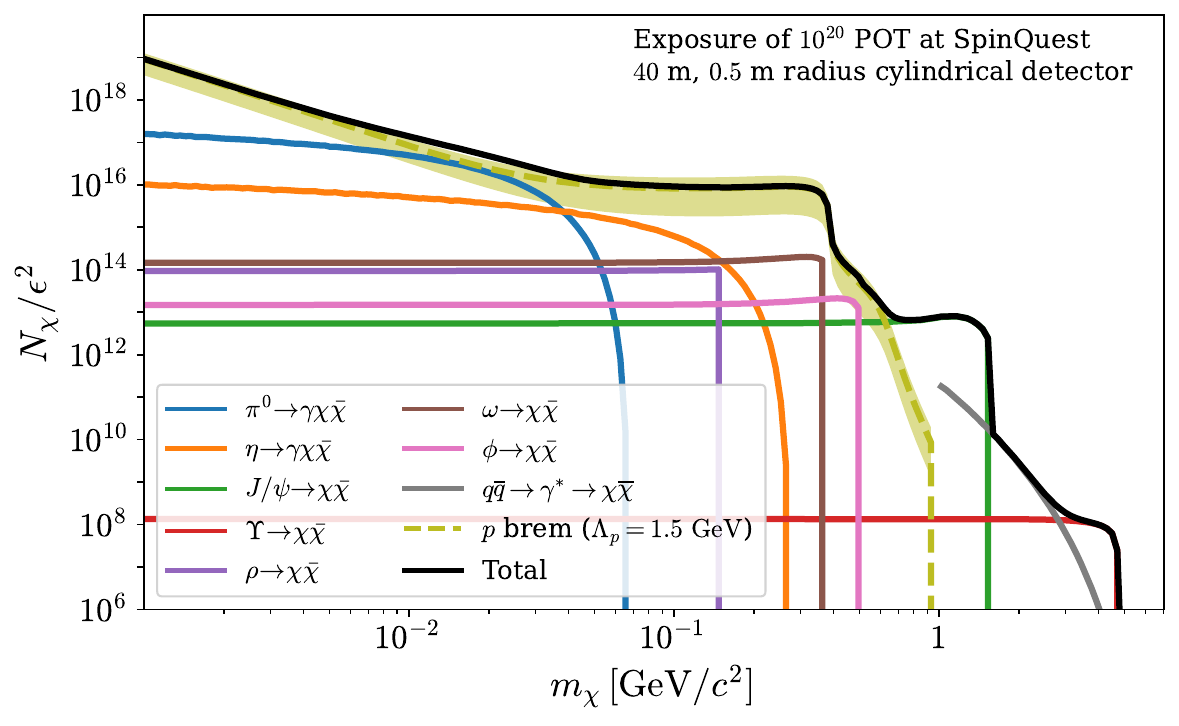}
        \caption{SpinQuest / LongQuest}
        \label{fig:spinquest_production}
    \end{subfigure}

    \vspace{1em}

    \begin{subfigure}{\linewidth}
        \centering
        \includegraphics[width=0.82\linewidth]{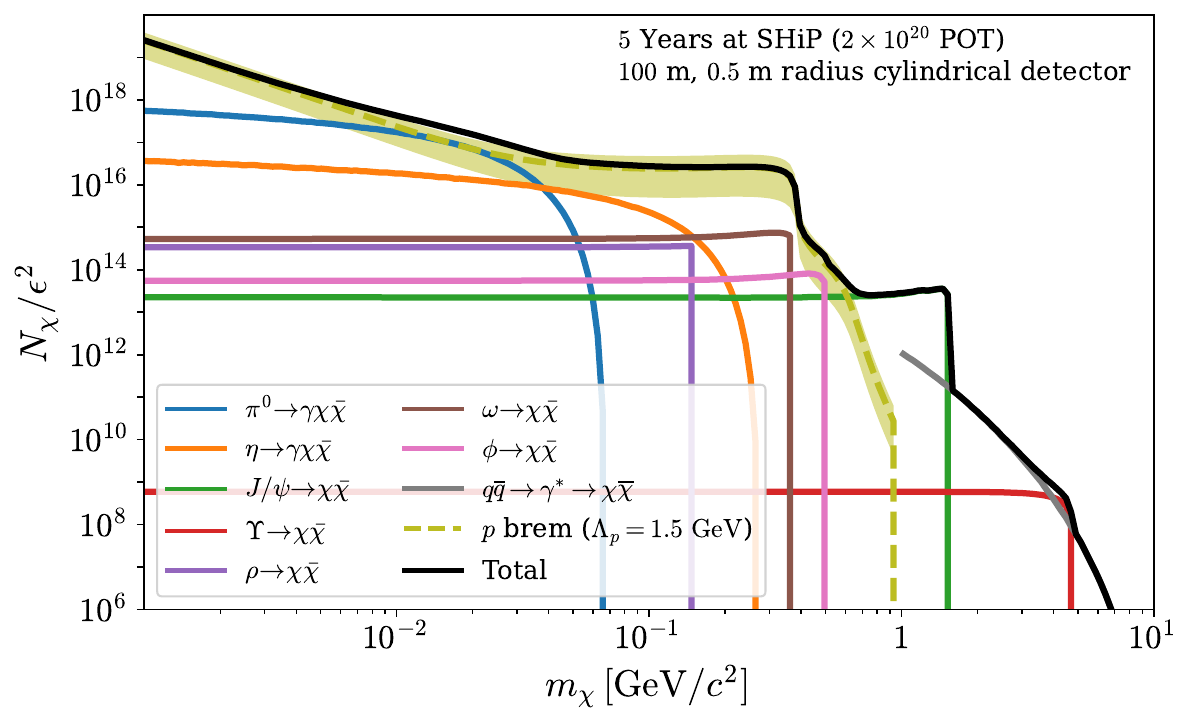}
        \caption{SHiP}
        \label{fig:ship_production}
    \end{subfigure}

    \caption{The mCP production projection for SpinQuest and SHiP from meson decays, Drell-Yan, and proton bremsstrahlung (sum of single tracks and double tracks hitting the same bar, with the Dirac electromagnetic form factor only). The uncertainty band corresponds to varying the associated cut-off scale $\Lambda_p \in [1,2]$ GeV, with the central value set to 1.5 GeV. The detector used in the simulation consists of $18\times 18$ scintillator bars/layer, located 40\,m (a) and 100\,m (b) downstream of the target. Proton bremsstrahlung dominates over all meson channels for $m_{\chi} < 0.5$ GeV.}
    \label{fig:mcp_production}
\end{figure}

So far, we have discussed the production of mCPs, for which the total yields is  $N_{\chi} = N_{\chi}^{\text{Meson}} + N_{\chi}^{\text{DY}}+N_{\chi}^{\text{Brem}}$.
The contributions of each production channel to $N_{\chi}$ are shown in Fig. \ref{fig:mcp_production}. mCPs entering the detector will interact with scintillator material and produce the photoelectrons, which are subsequently detected by PMTs.
The probability of observing mCP signals in $n$ layers follows a Poisson distribution, given by
\begin{align*}
    P = \qty(1-\exp[-\epsilon^2 N_{\text{PE}}])^n,
\end{align*}
where $N_{\text{PE}}$ is the average number of photoelectrons collected at a PMT when $\epsilon = 1$ and the additional factor of $\epsilon^2$ arises as the energy deposition scales with the square of the particle's electric charge.
We consider the identical configuration used in \cite{Tsai:2024wdh} for our setup and adopt the value $N_{\text{PE}} = 2.5 \times 10^5$ obtained from \texttt{GEANT4}~\cite{GEANT4:2002zbu} simulation.
The expected total number of signal events is given by $s = PN_{\chi}$.
We consider three-layer detectors ($n=3$) for both experimental facilities.

\section{Background Determination}\label{sec:backgrounds}

The backgrounds faced by the detector will be the same at SpinQuest and SHiP. The dominant sources are PMT dark rate, caused by the thermal emission of electrons from the photocathode (and other uncorrelated signals in timing and position through the detector), afterpulses from muons that travel through the detector, and deposits from shower particles that originate from both cosmogenic and beam muons. In this section, each source of background is considered for both SpinQuest and SHiP.

\subsection{Dark Rate}

As will be discussed in the following sections, this background source is expected to be dominant for both SpinQuest and SHiP. In PMTs, spurious ``dark current'' pulses can be produced by thermal electrons liberated from the photocathode. The contribution of this process to the background can be easily determined from the expected dark current of each PMT that comprises the detector. We assume a scenario of 100 Hz of dark rate (as used in Ref.~\cite{SUBMET}) for R7725 and predict the background rate using $n_{\rm bars\ per\ layer}\tau^{n_{\rm layers}-1}r^{n_{\rm layers}}$. 
The live time at both SpinQuest and SHiP is expected to be $\sim10^{6}$~s per year. For a three-layer detector of radius 0.5 cm, 324 bars/layer, this implies a background of 0.095 events per year for a dark rate of 100~Hz and 500~Hz, respectively. An overall background of 0.48 events is considered for both detectors from this process, assuming five years of data taking.

\subsection{Beam Muon Afterpulsing}

When beam muons traverse the detector, their deposits can be easily distinguished from signal by their much greater size. However, the PMTs that comprise the detector will experience afterpulsing that can increase the effective dark rate of each PMT by orders of magnitude for a short period. Tests on the bench and in-situ for the FORMOSA demonstrator experiment indicate a dead time of $\sim 1$~ms~\cite{Citron:2025kcy} is required to make this background negligible. At SHiP the muon flux is expected to be $\sim 0.2~\rm{Hz}/{\rm cm^2}$~\cite{Ferrillo:2023hhg}. Therefore, the muon rate is expected to be 5 Hz through each 25~$\rm{cm}^2$ bar face. This can be actively vetoed in the DAQ with a 1~ms deadtime after each muon, corresponding to only a $\sim0.5\%$ inefficiency. At SpinQuest, the detector will sit behind a large quantity of shielding material and so the muon flux is expected to be significantly lower than at SHiP. Therefore, the afterpulsing background will be negligible.

\subsection{Muon Showering}

Cosmogenic muons, as well as muons that originate in the beam dump, can cause an additional background if the muon produces showers in the cavern housing the detector or in the detector material. These shower particles can produce a background of small deposits in each layer if the muon itself does not travel through the detector. Such backgrounds were studied in detail for the milliQan detector in~\cite{milliQan:2021lne}. For cosmogenic muons a combination of active detector vetoes (a single deposit in each layer of the detector and with no hits in large panels that can be placed at both the front and back, as well as the sides of the detector to form a hermetic active veto), as well as a requirement that the deposits have a similar size in each layer of the detector, as expected for signal, are sufficient to make this background negligible. In addition, it should be possible to use information from the much larger SpinQuest and SHiP detectors to additionally veto deposits from showering beam muons that do not pass through the detector material. With the combination of these rejections, the muon showering background is expected to be negligible.

\section{Sensitivity Projections and Existing Constraints}\label{sec:sensitivity}

\begin{figure}
    \centering
    \includegraphics[width=1.0\linewidth]{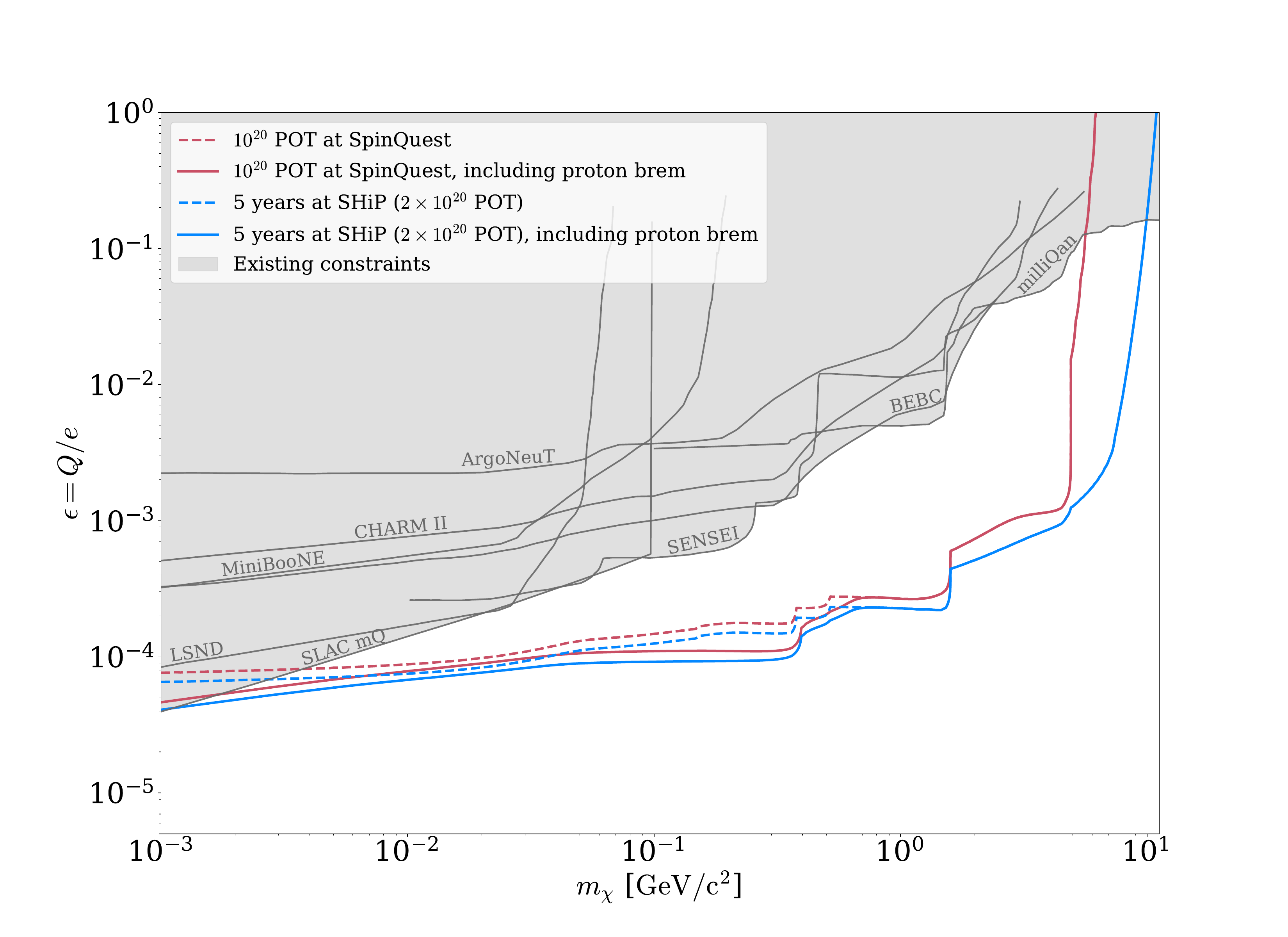}
    \caption{The expected 95\% CL limit for the proposed SHiP and SpinQuest mCP experiments. These results include contributions from 2-body and 3-body meson decays, Drell-Yan production and proton bremsstrahlung. Detector locations are assumed to be 40\,m for SpinQuest and 100\,m for SHiP downstream of the target. Results correspond to 5 years of operation in which $10^{20}$ POT and $2\times10^{20}$ POT are expected to be collected at SpinQuest and SHiP, respectively.}
    \label{fig:limit_plot_SHiP_DarkQuest}
\end{figure}

This section summarizes the projected sensitivity of the proposed mCP detectors at SHiP and SpinQuest and highlights the contributions of proton bremsstrahlung production. In Fig.~\ref{fig:limit_plot_SHiP_DarkQuest}, we show the projected 95\% CL exclusion limit under 5 years of nominal operation at SHiP and SpinQuest. Results correspond to an exposure of $2\times 10^{20}$ POT at SHiP and an accumulated exposure of $10^{20}$ POT at SpinQuest. The projections assume two identical three layer cylindrical detectors with a radius of 0.5 m, positioned 40 m downstream of the SpinQuest target and 100 m downstream of the SHiP target. Moreover, we assume a background of 0.5 events, which mainly originates from the PMT dark rate.

Including the proton bremsstrahlung production channel significantly enhances sensitivity in the $100~\text{MeV} \text{–} 1~\text{GeV}$ mass range (see Fig.~\ref{fig:limit_plot_SHiP_DarkQuest} for comparison with the meson baseline), with the expected limit on $\epsilon$ improving from $1.5\times 10^{-4}$ to $9.3\times 10^{-5}$ at $m_{\chi} \sim 200~\text{MeV}$ for SHiP. This channel produces a sharply forward-peaked flux of mCPs, yielding a substantially higher detectable rate than meson decays due to its strong collinear enhancement. The low-mass sensitivity gain originates from the rise in the mCP production rate via proton bremsstrahlung, which we explored using a factorized approach. See~\cref{app:phase_space_decompos} for details.

For the proton bremsstrahlung contribution of our analysis, we consider both single-track events and double-track events in which both mCPs traverse the same scintillator bar. An event is accepted if exactly one mCP hits a bar or if $\chi\bar{\chi}$ pair hits the same bar, and vetoed if the pair hits separate bars. This selection exploits the collinear nature of proton bremsstrahlung radiation, where the pair typically has a small opening angle, maximizing signal capture while reducing widely separated tracks.

Compared to current accelerator constraints, the combined channels yield significant sensitivity improvements. The enhanced limit is 3 times better in the low-mass region, exceeds 1 order of magnitude above $0.1$ GeV, and reaches up to 2 orders of magnitude in the $1$–$10$ GeV mass range. These projected results provide updated constraints for millicharged particles at SpinQuest and SHiP under fixed background assumptions of 0.5 events over 5 years. The complete simulation pipeline is open source and can be found in this~\href{https://github.com/exoticdarksectors/decaysimulation}{GitHub repository}.

\section{Discussion}\label{sec:discussion}
In this work, we propose dedicated mCP searches using scintillator-based detectors at high-intensity proton fixed-target facilities, with a focus on the forthcoming SpinQuest and SHiP experiments. In addition to meson decays and the Drell–Yan process, we include millicharged pair production via proton bremsstrahlung. We find that incorporating this contribution significantly increases the overall mCP yield, with a significant impact on the resulting experimental sensitivity. Moreover, the calculation method presented in~\cref{sec:pbrem} can be readily applied to other mCP search programs, offering a substantial opportunity to extend their reach.

For a scintillator detector, which is largely insensitive to the incoming mCP energy, higher intensity fixed-target beams are more advantageous than higher energy proton collisions at colliders for mCP masses below a few GeV. This is partly because the dominant proton bremsstrahlung yield, proportional to the inelastic $pA$ cross section, $\sigma_{pA}(s)$, only slowly increases with the center-of-mass energy, while benefiting from the nuclear target enhancement for (thick target) beam dump experiments. In addition, the shorter baselines of SpinQuest ($\sim40~\rm m$) and SHiP ($\sim100~\rm m$) provide a significantly higher geometric acceptance compared to the Forward Physics Facility (FPF) location ($\sim480 ~\rm m$), further enhancing the detectable mCP flux.

The modeling of proton bremsstrahlung production is an active area of ongoing study, as discussed in Sec.~\ref{sec:pbrem}. In this work, we simulate mCP pair production via initial-state radiation using the on-shell formalism of Refs.~\cite{Foroughi-Abari:2021zbm}. The dominant theoretical uncertainty arises from the treatment of the proton electromagnetic and off-shell form factors. Both analytic and data-driven approaches will be needed to reduce these uncertainties and further refine the on-shell framework, for example, see Refs.~\cite{Kling:2025udr,Allison:2025mom}. Additional production channels beyond initial-state radiation may also contribute, including final-state radiation in fully inelastic high energy collisions, which necessitates incorporating hadronization effects. Nevertheless, proton bremsstrahlung remains an important contribution and should be incorporated in accelerator-based and atmospheric mCP searches.

The focus of this work is on primary mCP production originating from the first interaction length in the target. Beyond the conventional production mechanisms considered here, secondary production from electromagnetic cascades~\cite{Zhou:2024aeu} inside thick beam-dump targets can be significant in the low-mass parameter space. In particular, resonant annihilation of positrons on atomic electrons, $e^+ e^- \to \gamma^\star(k^2)\, \gamma$ with subsequent decay $\gamma^\star(k^2) \to \chi \bar{\chi}$, can become effective at $k^2 \sim 2 m_e E_{e^+}$ and can contribute comparably to meson decays for $k \sim 10\text{--}50~\mathrm{MeV}$. Accounting for secondary production from electromagnetic cascades could further enhance the sensitivity reach at the small mass regime; nevertheless, it requires dedicated analysis.

The background prediction in our projections is estimated by drawing on previous studies and incorporating site-specific configurations.
More realistic assessments can be made with prototype in-situ measurements.
Hybrid detector designs that incorporate tracking or time-of-flight information, as already illustrated in the SpinQuest spectrometer~\cite{Apyan:2022tsd,Apyan:2025ldk} and in the SHiP timing detector~\cite{SHiP:2025ows},  can provide an additional handle that supports a negligible beam-related background assumption.
A detailed integration of the mCP detector with the existing infrastructure at these facilities will benefit from dedicated design studies in the future.

The two facilities we focus on in this work—SpinQuest and SHiP—offer complementary strengths for mCP searches.
Both sites provide substantial shielding, which is advantageous for both beam-related and cosmogenic background mitigation.
Compared to SHiP, the SpinQuest location also benefits from a detector located closer to the target, leading to an enhanced flux of mCPs.
In contrast, SHiP's $400$ GeV proton beam yields larger Drell-Yan production for $m_{\chi} \gtrsim 1$ GeV, thus providing better sensitivity in the high-mass region. Taken together, our findings support a positive outlook that the next generation of intensity-frontier experiments, including SpinQuest and SHiP, will deliver substantial gains in the discovery potential for millicharged particles.

\section*{Acknowledgement}
We thank Adam Ritz, Felix Kling, and Peter Reimitz for useful discussions. YC and FL are
supported by the US Department of Energy under award number DE-SC0008541. SF is supported by a Subatomic Physics Discovery Grant (individual) from the Natural Sciences and Engineering Research Council of Canada. This research was enabled in part by support provided by the Digital Research Alliance of Canada (\url{https://alliancecan.ca}). MC is supported by the DOE Office of Science under award number SC0009999.
This research is partially supported by Los Alamos National Laboratory's Laboratory Directed Research and
Development (LDRD) program. YDT thanks the generous support from the LANL Director's Fellowship.
This research was supported in part by a grant NSF PHY-2309135 to the Kavli Institute for Theoretical Physics (KITP).
This work was partially performed at the Aspen Center for Physics, supported by National Science Foundation grant No.~PHY-2210452. This research was partly supported by the NSF under Grant No.~NSF PHY-1748958.  This work was partially supported by the National Research Foundation of Korea (NRF) grant funded by the Korean government (MSIT) (RS-2021-NR059935 and RS-2025-00560964) and a Korea University grant.

\newpage
\appendix
\crefalias{section}{appendix}

\section{Millicharged Pair Production via Proton Bremsstrahlung}\label{app:pBrem}

In this appendix, we compute the cross section of the forward production of mCP pairs via proton bremsstrahlung,
\begin{equation}\label{eq:app-brem-scattering}
p(p)+p(p_t)\ \rightarrow\ \chi(p_\chi)+\bar{\chi}(p_{\bar{\chi}})+f(p_f),
\end{equation}
where a mCP pair is emitted from the incoming beam proton $p$, which then undergoes inelastic scattering with the target proton $p_t$, and \(f\) denotes the inclusive hadronic final state. The virtual photon carries momentum \(k = (p_\chi+p_{\bar{\chi}})\) with invariant mass squared \(k^2 \ge 4m_\chi^2\). 
We adopt the quasi-real approximation (QRA) for initial-state radiation (ISR) in inelastic scattering as developed in Ref.~\cite{Foroughi-Abari:2021zbm}. In this approach, the intermediate proton $p^\prime$ in the ISR diagram in Fig.~\ref{fig:FeymCP} is treated as nearly on shell, allowing the process to factorize into the emission of the $\chi\bar{\chi}$ pair and the subsequent hard proton–proton interaction. The emission vertex consists of the dark state pair current
\begin{equation}\label{eq:paircurrent}
\mathcal{M}_{\chi}^{\nu} = \epsilon e \bar{u}(p_\chi)\gamma^{\nu}v(p_{\bar\chi}),
\end{equation}
and the proton electromagnetic vertex
\begin{align}
\mathcal{M}_{\rm splt}^\mu
= (ie)\,\bar{u}^{s^\prime}(p^\prime)
\left[
\gamma^\mu F_1^p(k^2)
+ \frac{i\sigma^{\mu\nu}(p^\prime-p)_\nu}{2m_p}F_2^p(k^2)
\right]
u^{s}(p),
\end{align}
where $t=k^2$ is timelike at the ISR vertex, and both the Dirac $F_1^p(t)$ and Pauli $F_2^p(t)$ proton electromagnetic form factors are retained. The parametrization of nucleon electromagnetic form factors in the timelike region, along with comparisons and fits to data, is discussed in~\cite{Foroughi-Abari:2024xlj,Kling:2025udr}. The outgoing intermediate proton is taken to be on-shell, and the energy conservation at the $pp^\prime\gamma^\ast$ vertex is not enforced, so that $p'\neq p-k$. Instead, the proton recoil follows from momentum conservation, and the intermediate proton's energy is given by
\begin{equation}
E_{p^\prime}
=
\sqrt{\big(\vec p - \vec p_\chi - \vec p_{\bar\chi}\big)^2 + m_p^2}~.
\end{equation}

Under QRA factorization, the intermediate proton propagator can be replaced by the polarization sum for an on-shell proton, so then the amplitude for the process~(\ref{eq:app-brem-scattering}) can be approximated as
\begin{align}\label{eq:bremAmp}
\mathcal{M}_{pp \rightarrow \chi\bar{\chi}\,f}
\simeq
\frac{
\sum_{s^\prime}
\mathcal{M}_{p^\prime p \rightarrow f}
\,\mathcal{M}_{\rm splt}^{\mu}
}{
(p-k)^2 - m_p^2
}
\;
\frac{-ig_{\mu\nu}}{k^2}
\;
\mathcal{M}_{\chi}^\nu,
\end{align}
where the sum runs over helicities of the intermediate proton.

The differential cross section for the process is
\begin{equation}
\mathrm{d}\sigma
=
\frac{
\left|\overline{\mathcal{M}}\right|^2
}{
4E_p E_{p_t}\,
|\vec v_p - \vec v_{p_t}|
}
\,
\mathrm{d}\Phi,
\end{equation}
with the phase space element
\begin{align}
d\Phi = 
\frac{\mathrm{d}^3 p_\chi}{(2\pi)^3 2E_\chi}
\frac{\mathrm{d}^3 p_{\bar\chi}}{(2\pi)^3 2E_{\bar\chi}}
\prod_f \frac{\mathrm{d}^3 p_f}{(2\pi)^3 2E_f}
(2\pi)^4
\delta^{(4)}
\!\left(
p + p_t - p_\chi - p_{\bar\chi} - \sum_f p_f
\right).
\label{eqn:phase_space_QRA}
\end{align}

After integrating over the final-state system $f$, the cross section factorizes into the splitting probability and the $pp$ scattering cross section evaluated at reduced energy with $s^\prime\simeq (p-k+p_t)^2$, which can be expressed as
\begin{align}\label{eqn:QRA_XS}
d\sigma^{pp \rightarrow \chi\bar{\chi}\,f}
&\simeq
\frac{\left|\overline{\mathcal{M}^{\,p\rightarrow p^\prime \chi \bar{\chi}}}\right|^2}{\left[(p-k)^2 - m_p^2\right]^2}
\frac{|\vec p - \vec k|}{|\vec p|} \frac{\mathrm{d}^3 p_\chi}{(2\pi)^3 2E_\chi}
\frac{\mathrm{d}^3 p_{\bar\chi}}{(2\pi)^3 2E_{\bar\chi}}
\; \\
& \quad \times \int \frac{1}{4E_{p^{\prime}} E_{p_t}|\vec v_{p^\prime} {-} \vec v_{p_t}|}\prod_f \frac{\mathrm{d}^3p_f}{(2\pi)^3 2E_f}|\overline{\mathcal{M}^{p^{\prime}p\rightarrow f}}|^2 (2\pi)^4\delta(p^{\prime}{+}p_{t}{-} p_f)   \nonumber
\\[2pt]
&\equiv
\mathrm{d}\mathcal{P}_{p \rightarrow p^\prime\chi\bar\chi}
\times
\sigma^{pp \rightarrow f}(s^\prime),
\end{align}
corresponding to the splitting probability of Eq. (\ref{eq:master}), with the overall matrix element for the splitting and emission
\begin{equation}\label{eq:MEemission}
    \left|\overline{\mathcal{M}^{\,p\rightarrow p^\prime \chi \bar{\chi}}}\right|^2 \equiv  \epsilon^2 (4\pi\alpha)^2
\frac{S^{\mu\nu}\chi_{\mu\nu}}{k^4}.
\end{equation}
In the argument for energy conservation in the delta function in~\cref{eqn:phase_space_QRA}, we also set $p-k\simeq p^\prime$. This difference is negligible when the intermediate proton is near on-shell for low-$p_T$ emission. The three-body phase space $\{p^\prime,p_\chi,p_{\bar\chi}\}$ has five independent variables after removing three from momentum conservation and one by axial symmetry around $\vec{p}$.

The dark current tensor is captured in the final state
spin-summed matrix element $\chi_{\mu\nu}$ given by
\begin{equation}
\chi_{\mu\nu}
=
\mathrm{Tr}\!\left[
(\slashed p_\chi + m_{\chi})
\gamma_{\mu}
(\slashed p_{\bar\chi} - m_{\chi})
\gamma_{\nu}
\right].
\end{equation}
The proton splitting structure, averaged over initial and summed over final spins, is described by the terms
\begin{align}
S^{\mu\nu} &= \frac{1}{2}\,
\mathrm{Tr}\!\left[(\slashed p^{\prime}{+}m_p)\gamma^\mu(\slashed p{+}m_p)\gamma^\nu\right]\,\big|F_1(k^2,p^{\prime 2})\big|^2 \\ \nonumber
&\;\;\; 
 + \frac{1}{2}\,
\mathrm{Tr}\!\left[(\slashed p^{\prime}{+}m_p)\sigma^{\mu\alpha}(\slashed p{+}m_p)\sigma^{\nu\beta}\right]\frac{(p{-}p^\prime)_\alpha (p{-}p^\prime)_\beta}{4m_p^2}\,\big|F_2(k^2,p^{\prime 2})\big|^2 \\ \nonumber
&\;\;\; 
 + \frac{i}{2}\,
\mathrm{Tr}\!\left[(\slashed p^{\prime}{+}m_p)\gamma^\mu(\slashed p{+}m_p)\sigma^{\nu\alpha}\right]\frac{(p{-}p^\prime)_\alpha}{2m_p}\,F_1(k^2,p'^2)F_2^\star(k^2,p'^2) \\ \nonumber
&\;\;\; 
 - \frac{i}{2}\,
\mathrm{Tr}\!\left[(\slashed p{+}m_p)\gamma^\nu(\slashed p^{\prime}{+}m_p)\sigma^{\mu\alpha}\right]\frac{(p{-}p^\prime)_\alpha}{2m_p}\,F_1^\star(k^2,p'^2)F_2(k^2,p'^2) .
\end{align}
Note that $k_\mu \chi^{\mu\nu}=0$ due to the Ward identity for the intermediate photon.

The matrix element~\cref{eq:MEemission}, including the form factor reduces to
\begin{align}\label{eq:FFcontributions}
S^{\mu\nu}\chi_{\mu\nu} = \mathcal{W}_{1}\big|F_1(k^2,p^{\prime 2})\big|^2 + \mathcal{W}_{2}\big|F_2(k^2,p^{\prime 2})\big|^2 +\mathcal{W}_{12}\mathrm{Re}\left[F_1(k^2,p^{\prime 2})F_2^\star(k^2,p^{\prime 2})\right] 
\end{align}
with 

\begin{align}
\mathcal{W}_{1} = & \, 16 \left[m_\chi^2 (p{-}p')^2 - m_p^2k^2 +2(p {\cdot} p_{\bar{\chi}})(p' {\cdot} p_{\chi}) + 2 (p' {\cdot} p_{\bar{\chi}})(p {\cdot} p_{\chi})\right],
\\ \nonumber
\mathcal{W}_{2} = & \, 16\left[\left(p_\chi {\cdot}\left(p{-}p'\right)\right)\left(p_{\bar{\chi}} {\cdot} \left(p{-}p'\right)\right)\right] - \frac{(p{-}p')^2}{m_p^2} \left[k^2\left(p{-}p'\right)^2 + 4\left(p_\chi {\cdot}\left(p{+}p'\right)\right)\left(p_{\bar{\chi}} {\cdot} \left(p{+}p'\right)\right) +16m_\chi^2m_p^2 \right],
\\ \nonumber
\mathcal{W}_{12} = & \, 4\left[ 4\left(p_\chi {\cdot}\left(p{-}p'\right)\right)\left(p_{\bar{\chi}} {\cdot} \left(p{-}p'\right)\right) - \left(p{-}p'\right)^2 \left( k^2 {+} 4m_\chi^2 \right) \right].
\end{align}

\begin{figure}[t]
\includegraphics[width=0.80\textwidth]{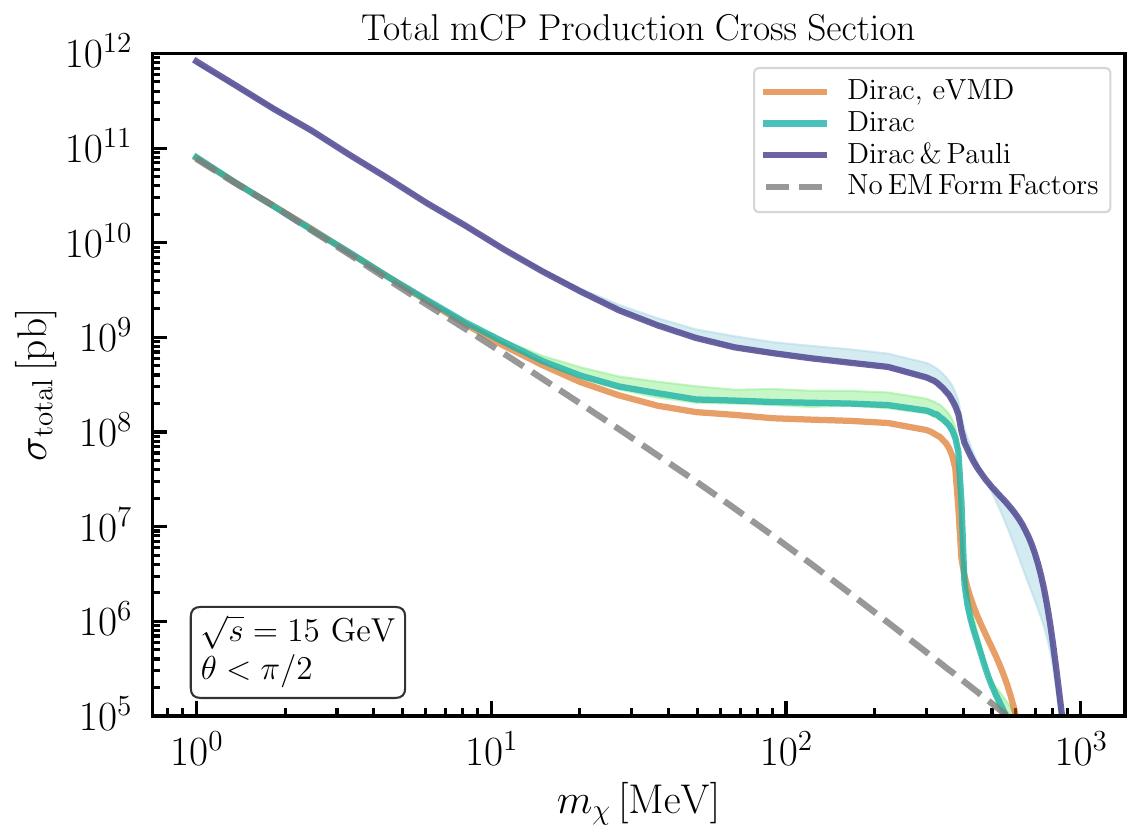}
\caption{Comparing total mCP production rates via proton bremsstrahlung as a function of mCP mass, including Dirac-only (green) and combined Dirac–Pauli (blue) form factors with uncertainties from model variations~\cite{Kling:2025udr}, Dirac-only eVMD~\cite{Faessler:2009tn} (orange), and no form factor (dashed gray), for the full forward-hemisphere phase space.}
\label{fig:brem_total_formfactors}
\end{figure}

In~\cref{fig:brem_total_formfactors}, we compare the impact of nucleon electromagnetic form factor contributions, as in~\cref{eq:FFcontributions}, on the total mCP production cross section, corresponding to the full phase space allowed by forward-hemisphere angular cuts. This includes the case with both Dirac and Pauli form factor contributions, where the electromagnetic form factors are taken from the improved treatments presented in Ref.~\cite{Kling:2025udr}. The associated uncertainty band reflects differences between alternative fitting models for the nucleon electromagnetic structure. We also consider predictions using only the Dirac form factor within the extended Vector Meson Dominance (eVMD) approach~\cite{Faessler:2009tn}, as well as the production yield in the absence of any form factor.

\section{mCP Phase Space Decomposition}\label{app:phase_space_decompos}

In this appendix, we explore the origin of the observed rise in the millicharged particle production cross section via proton bremsstrahlung at low masses. Similar behavior has been noted in electromagnetic production of dark states through lepton bremsstrahlung in QED-like processes~\cite{Freytsis:2009bh,Bjorken:2009mm,Izaguirre:2014bca,Gninenko:2018ter,Zhou:2024aeu,Blinov:2025aha}, where the total rate shows an approximate $1/m_\chi^2$ scaling for small $m_\chi$. This scaling arises from the interplay between the photon propagator, which contributes $1/k^4 \sim 1/m_\chi^4$ when the virtuality is set by the pair mass, and the two-body phase space, which provides a compensating factor of $m_\chi^2$.

We begin by deriving an approximate expression for the total mCP pair production rate, following~\cite{Beranek:2013yqa, Gninenko:2018ter, Chu:2018qrm}. The cross section for the process~\cref{eq:brem-scattering} can be factorized further by the phase space decomposition
 
\begin{equation}
    \mathrm{d} \Phi_{pp \rightarrow \chi \bar{\chi}+f}=\frac{\mathrm{~d} k^2}{2 \pi} \mathrm{~d} \Phi_{pp \rightarrow \gamma^\star + f}  \mathrm{~d} \Phi_{\gamma^\star \rightarrow \chi \bar{\chi}},
\end{equation}
where the intermediate photon can be treated as on-shell with invariant mass $k^2$. This factorization is exact and follows from inserting an additional integral together with an appropriate $\delta^{(4)}$ constraint that accounts for the extra integration over $k^2$.

Following Ref.~\cite{Uhlemann:2008pm}, the absolute square of the amplitude in~\cref{eq:bremAmp}, summed over the spins and polarizations of the outgoing states and averaged over the initial proton spins with the denominator of the photon propagator suppressed, can also be factorized by
\begin{align}\label{eq:MEdecomposed}
\left|\overline{\mathcal{M}_{pp \rightarrow \chi \bar{\chi}+f}}\right|^2
&=\frac{1}{4}\sum_{\rm{spins/pols.}}\left|\mathcal{M}^\mu\left(-ig_{\mu \nu}\right) \mathcal{M}_{\chi}^\nu\right|^2
\\ \nonumber
& \rightarrow \frac{1}{4}\sum_{\rm{spins/pols.}}\left|\mathcal{M}^\mu\left(-g_{\mu \nu}+\frac{k_\mu k_\nu}{k^2}\right) \mathcal{M}_{\chi}^\nu\right|^2
\\\nonumber
&\simeq \frac{1}{4}\sum_{\rm{spins/pols.}}\sum_{\lambda_1}\left|\mathcal{M}^\mu \epsilon_\mu^{\lambda_1}(k)\right|^2 \times \frac{1}{3} \sum_{\lambda_2}\left|\mathcal{M}_{\chi}^\nu \epsilon_\nu^{\star\lambda_2}(k)\right|^2
\\ \nonumber
&= \left|\overline{\mathcal{M}_{pp \rightarrow\gamma^\star +f}}\right|^2 \times \left|\overline{\mathcal{M}_{\gamma^\star\rightarrow \chi \bar{\chi}}}\right|^2 \,.
\end{align}
Here, $\mathcal{M}_{\chi}^\nu$ denotes the spinor bilinear associated with the $\chi\bar{\chi}$ pair, as defined in~\cref{eq:paircurrent}, and $\mathcal{M}^\mu$ corresponds to the remaining amplitude of the $pp$ subprocess. In the second line we shift the numerator of the intermediate photon propagator, and in the third line we rewrite the expression in terms of the polarization sum of a massive vector boson with invariant mass $k^2$, with $\sum \epsilon^{\star \lambda}_\mu(k) \epsilon_\nu^\lambda(k) =-g_{\mu\nu} + k_\mu k_\nu/k^2$. After rearranging the polarization sums and neglecting polarization correlations, the expression fully factorizes into the production of a virtual photon and its subsequent decay. This rearrangement is exact for on-shell intermediate states when only total decay rates are of interest. However, spin correlations between production and decay can modify the resulting energy and angular distributions of the decay products~\cite{Feng:2025gji}.

Note that the Ward identity for the millicharged current, i.e., $k_\mu\mathcal{M}_\chi^\mu=k_\mu\mathcal{M}^\mu=0$, ensures that the $k_\mu k_\nu$ term in the second line vanishes for external on-shell fermions. Consequently, any modification to the polarization sum in~\cref{eq:MEdecomposed} must remain consistent with this requirement.

\begin{figure}[htb]
\includegraphics[width=0.80\textwidth]{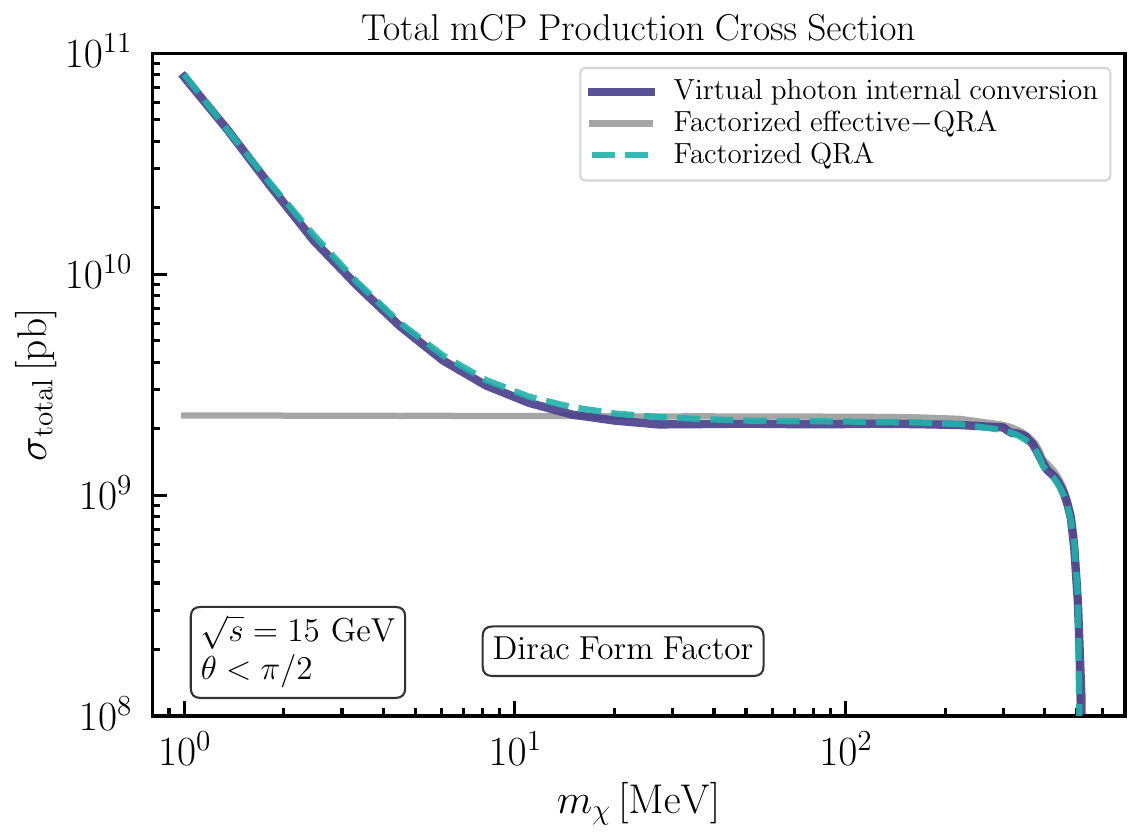}
\caption{Total millicharged particle production cross section via proton bremsstrahlung, integrated over the forward hemisphere ($\theta < \pi/2$). We compare the full numerical calculation (blue), which keeps the virtual photon propagator intact, with the factorized expressions. The expected low-mass scaling agrees with the factorized formula using the QRA splitting function for massive on-shell photons (dashed). At very low mass, the factorized formula, combined with the effective-QRA (Dawson corrected), flattens (gray). Only the proton Dirac electromagnetic form factor at the proton splitting vertex is included. See text for details.}
\label{fig:brem_total_factorized}
\end{figure}

Combining all pieces, the millicharged particle production cross section can be expressed in the following factorized form
\begin{align}\label{eq:XStotaldecomposed}    
\mathrm{~d}\sigma(pp \rightarrow \chi\bar{\chi} + f) 
&=\frac{\mathrm{d} k^2}{2 \pi} \frac{1}{k^4} \mathrm{~d} \Phi_{pp \rightarrow \gamma^\star + f} \left|\overline{\mathcal{M}_{pp \rightarrow \gamma^\star + f}}\right|^2\left(\int \mathrm{d} \Phi_{\gamma^\star \rightarrow \chi \bar{\chi}}\left|\overline{\mathcal{M}_{\gamma^\star \rightarrow \chi \bar{\chi}}}\right|^2\right) \\ \nonumber
& = \frac{\mathrm{d} k^2}{2 \pi} \frac{1}{k^4} \mathrm{~d} \sigma(pp \rightarrow \gamma^\star + f) \times 2 k \Gamma(\gamma^\star \rightarrow \chi\bar{\chi}),
\end{align}
where the decay width of a vector boson with virtual mass squared $k^2$ into millicharged pairs is given by

\begin{equation}
    \Gamma(\gamma^\star \rightarrow \chi\bar{\chi})= \frac{\epsilon^2 \alpha}{3} k \sqrt{1-\frac{4 m_\chi^2}{k^2}}\left(1+\frac{2 m_\chi^2}{k^2}\right),
\end{equation} 
and the directions of $\chi$-particles have been integrated out.

For the production of a virtual photon with invariant mass $k^2$, as part of the decomposition formula in~\cref{eq:XStotaldecomposed}, we employ the quasi-real approximation splitting kernel~\cite{Foroughi-Abari:2021zbm},
\begin{align}\label{eq:diffSplit}
    \mathrm{~d}\sigma^{pp\rightarrow \gamma^\star + f}(s) \approx w_{\rm QRA}(z,p_T^2)\mathrm{~d}z\mathrm{~d}p_T^2\times\sigma_{pp}^{\rm NSD}(s^{\prime}),
\end{align}
with 
\begin{align}\label{eq:SplitVectorMain}
w_{\rm QRA}(z,p_T^2) &= \frac{\alpha}{2\pi H}
\bigg[z-z(1{-}z) \Big(\frac{2m_p^2{+}k^2}{H}\Big) + \frac{H}{2z k^2} \bigg] \big|F^p_1(k^2,p'^2)\big|^2, 
\end{align}
where we restrict to the Dirac form factor for now, and $H(z,p_T^2)\equiv p_T^2{+}z^2m_p^2{+}(1{-}z)k^2$. We do not include the Dawson corrected splitting function~\cite{Foroughi-Abari:2024xlj}, which modifies the transverse polarization of the dark photon and has recently been applied to millicharged particle production in Ref.~\cite{Kling:2025udr}.

For sufficiently small \(m_\chi\), the dominant contribution to massive photon production arises from the last term in the bracket of~\cref{eq:SplitVectorMain}, which is proportional to \(1/k^{2}\). An approximate expression for the total \(\chi\bar{\chi}\) production rate in the low-mass regime can be obtained by integrating \cref{eq:XStotaldecomposed} over the invariant mass \(k^{2}\)
\begin{equation}\label{eq:brem_mass_scaling}
   \sigma(pp \rightarrow \chi\bar{\chi} + f)
   \approx
   \frac{\epsilon^{2}\alpha^{2}}{12\pi^{2}}
   \frac{1}{(2m_\chi)^{2}}
   \int_{z_{\rm min}}^{z_{\rm max}} \frac{dz}{z}
   \int_{0}^{p_{T,\rm max}^{2}} dp_T^{2}
   \int_{1}^{\infty} \frac{dy}{y^{2}}
   \sqrt{1 - \frac{1}{y}}
   \left(1 + \frac{1}{2y}\right),
\end{equation}
where \(y \equiv k^{2}/(4m_\chi^{2})\). The explicit \(1/m_\chi^{2}\) scaling in this expression is consistent with the qualitative features of \(\chi\bar{\chi}\) production mediated by a virtual photon in quasi-elastic QED-like processes, e.g., $e^{\pm}/\mu^{\pm}N\rightarrow e^{\pm}/\mu^{\pm}N +\chi\bar{\chi}$~\cite{Bjorken:2009mm,Izaguirre:2014bca,Chu:2018qrm}.

In~\cref{fig:brem_total_factorized}, we compare the canonical calculation performed by keeping the virtual photon propagator intact as in~\cref{eqn:QRA_XS}, including matrix element~\cref{eq:MEemission}, with the factorized expression derived in~\cref{eq:XStotaldecomposed}. The rate from virtual-photon internal conversion computation follows the expected scaling in the low-mass regime and agrees with the approximate factorized formula when the QRA splitting function of Ref.~\cite{Foroughi-Abari:2021zbm} for a massive vector boson is used. This factorized formula disagrees with that of Ref.~\cite{Kling:2025udr} due to a different polarization sum applied at the splitting vertex, which leads to the rise in the production rate in the low-mass regime not being properly captured by the effective QRA splitting function with the Dawson subtraction~\cite{Foroughi-Abari:2024xlj}.

\bibliography{references}

\end{document}